# On the existence of a standard rod in the Universe

I.N. Pashchenko[1], B.V. Komberg[2]
*Astro Space Center, Lebedev Physical Institute of Russian Academy of Science, 84/32 Profsoyuznaya st., Moscow, 117997, Russia*

**ABSTRACT**
Using standard cosmological model we show that the cores of ultra-compact radio sources observed with ground-based Very Long Baseline Interferometry (VLBI) on the angular scales of milliarcseconds cannot be used as a reasonable standard unit of linear size. "Luminosity – linear size" correlation obtained by many authors for ultra compact radio sources has different origin than that for the radio galaxies and quasars on the angular scales of arcminutes. It is just the manifestation of the fact that ground-based VLBI networks are unable to resolve VLBI-cores and the Malmquist bias presents. Thus, the cores of compact radio sources can't be used as "standard rods" at least with resolution offered by ground-based VLBI. This conclusion is illustrated on 15 GHz VLBA sample of radio sources.

## 1. INTRODUCTION

The idea of using the "angular size – redshift" relation for testing cosmological models has been known for many years. The main drawback of this method is that one needs the standard rod – some observable linear scale that is free of evolution with redshift - which is however hard to find. In radio sky the first obvious candidates – the radio loud AGN (radiogalaxies, quasars, blazars) with linear radio structures – jets, observed on scales of arcseconds-arcminutes, failed to give us such scale. Namely, angular size dependence of linear radio structures with redshift has shown "Euclidean" 1/z behavior in contrast with the prediction of Friedman-Lemaitre-Robertson-Walker (FLRW) models without evolution (Kapahi, 1989). Authors of (Singal, 1993) and (Nilsson et al., 1993) considered this contradiction as a result of "linear-size - luminosity" correlation. More luminous sources (which are observed mainly at higher redshifts in flux limited samples) possess the smaller linear sizes. Authors of (Blundell et al., 1999) called this effect the "youth-redshift degeneracy" and linked the smaller linear sizes of powerful radio sources with them being in the earlier stage of evolution.

There are ultra compact radio sources among radio loud AGN that can be observed with VLBI on the scales of milliarcseconds. The use of these objects in the context of "standard rods" is justified by several reasons (Kellerman, 1993, Gurvits et al., 1999). First of all, with lifetimes of years (not tens of millions years as radio structures on scales of arcseconds) this objects may be free of some evolutional effects that could confuse interpretation of the cosmological test. Also flux limited samples should lie within a narrow range of projection angles thus minimizing projection effects (Vermeulen, Cohen, 1994). And finally as the origin of the radio activity may be connected with the properties of "central engine" – mass or spin of the super massive black hole (SMBH) in the center of host galaxy and the latter do not seem to differ significantly among radio loud AGN. The properties of compact radio sources may be enclosed in some narrow ranges to treat compact radio sources as "standard rods" (Gurvits et al., 1999).

[1] E-mail: in4pashchenko@gmail.com
[2] E-mail: bkomberg@asc.rssi.ru

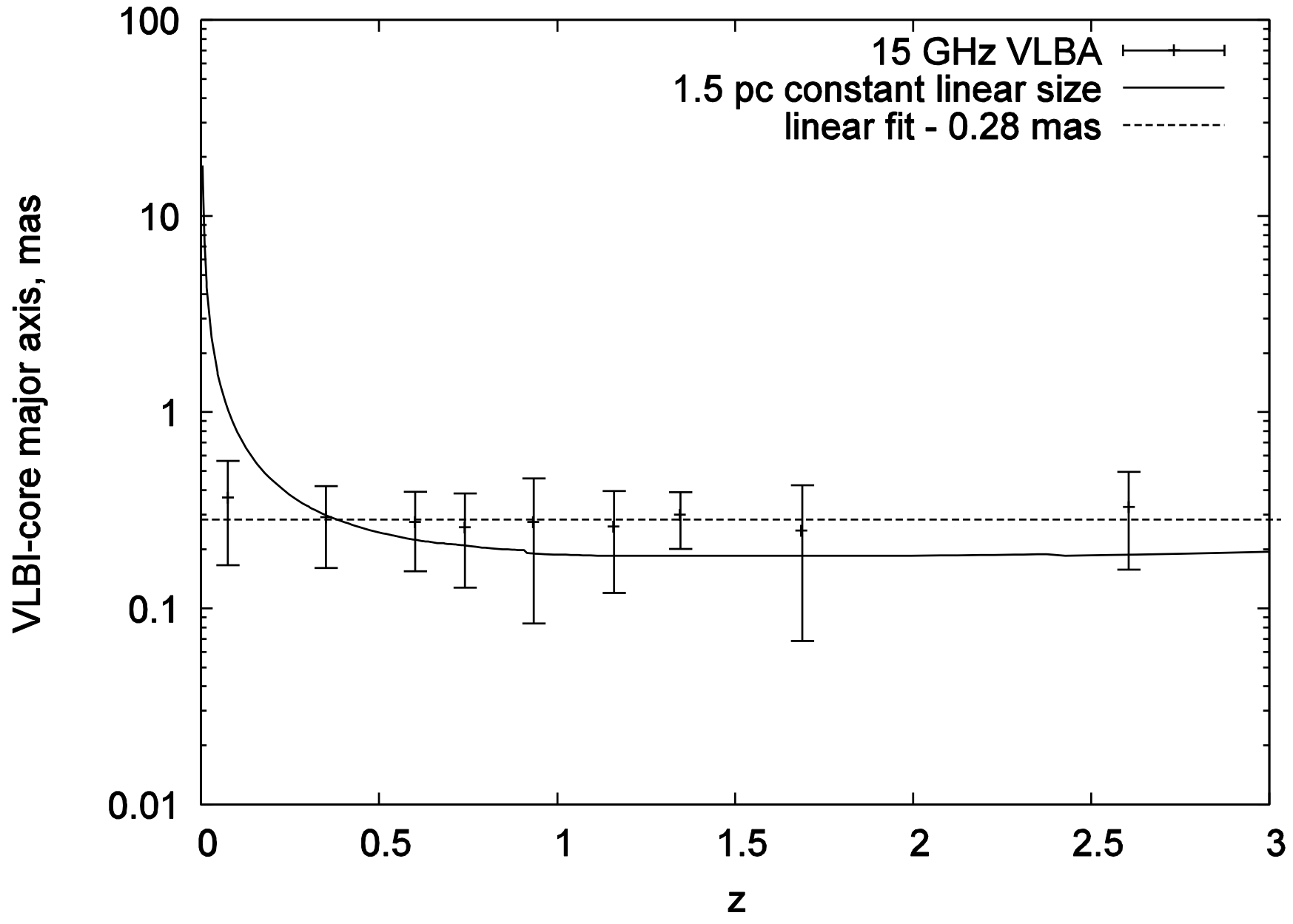

Fig. 1 Angular size - redshift diagram for 186 compact sources binned in redshift where VLBI-core major axis size is used as a characteristic angular size for each source. Dashed line represents standard rod of 1.5 pc (~$1h^{-1}$pc). Solid line is the weighted linear fit to binned data points

Nevertheless, recent claims that compact radio sources provide standard rods (Jackson, 2008) need to be clarified. In this paper we discuss the possibility to use the compact sources and, especially, their most compact and bright components – VLBI-cores, as standard rods.

## 2. COMPACT SOURCES AS STANDARD RODS

Compact radio sources typically consist of a well defined "core" (also called "VLBI-core") – often the brightest detail on radio map with flattest spectra that is almost always the base of a nearly linear structure with steeper spectra fading out of the core – the jet (VLBI-jet). Jet consists of several components or "knots" – bright spots on the overall smooth intensity distribution, or entirely consists of such "knots". Actually appearance of jet on VLBI map depends on the signal-to-noise ratio achieved in the VLBI experiment. Jets of some AGN on milliarcsecond scales can appear as almost "knotless", being dominated by smooth "diffusion" radio emission.

Nontrivial morphology of compact radio sources raises the question of what should be used as the characteristic angular scale for individual sources? Some works use separation between core and the most distant component in the jet brighter than some threshold in flux as a measure of angular scale. In his pioneer work (Kellermann, 1993) used such data for 79 objects, observed on VLBI at 5 GHz. After binning in the redshift, the author has shown that these results were consistent with standard FLRW cosmological model. Also it is worth to mention the work of (Gurvits, 1994), who used a large (337 sources) VLBI data compilation from 2.3GHz survey (Preston et al., 1985). The author used ratio of visibility amplitudes at two points of uv-plane (compactness) as a measure of characteristic angular size. He ignored sources with $z$<0.5 justifying this by apparent "linear size - luminosity" correlation which will be discussed below. Doing so, he found the data to be consistent with low-density FLRW-model.

In principle the angular size of VLBI-core (which is often modeled as elliptical Gaussian with 3 parameters: major, minor axis and positional angle) could also be used as characteristic angular scale. Author of (Jackson, 2008) used a size of VLBI-core major axis on 5GHz as a characteristic angular scale for 271 sources to consider apparent size – redshift relation. He discusses well known "linear size - luminosity" correlation and concludes that compact radio sources at $z$>0.5 do comprise a standard rods. Below we will see that cores of compact radio sources can't be used as standard rods at least with resolution offered by ground based VLBI.

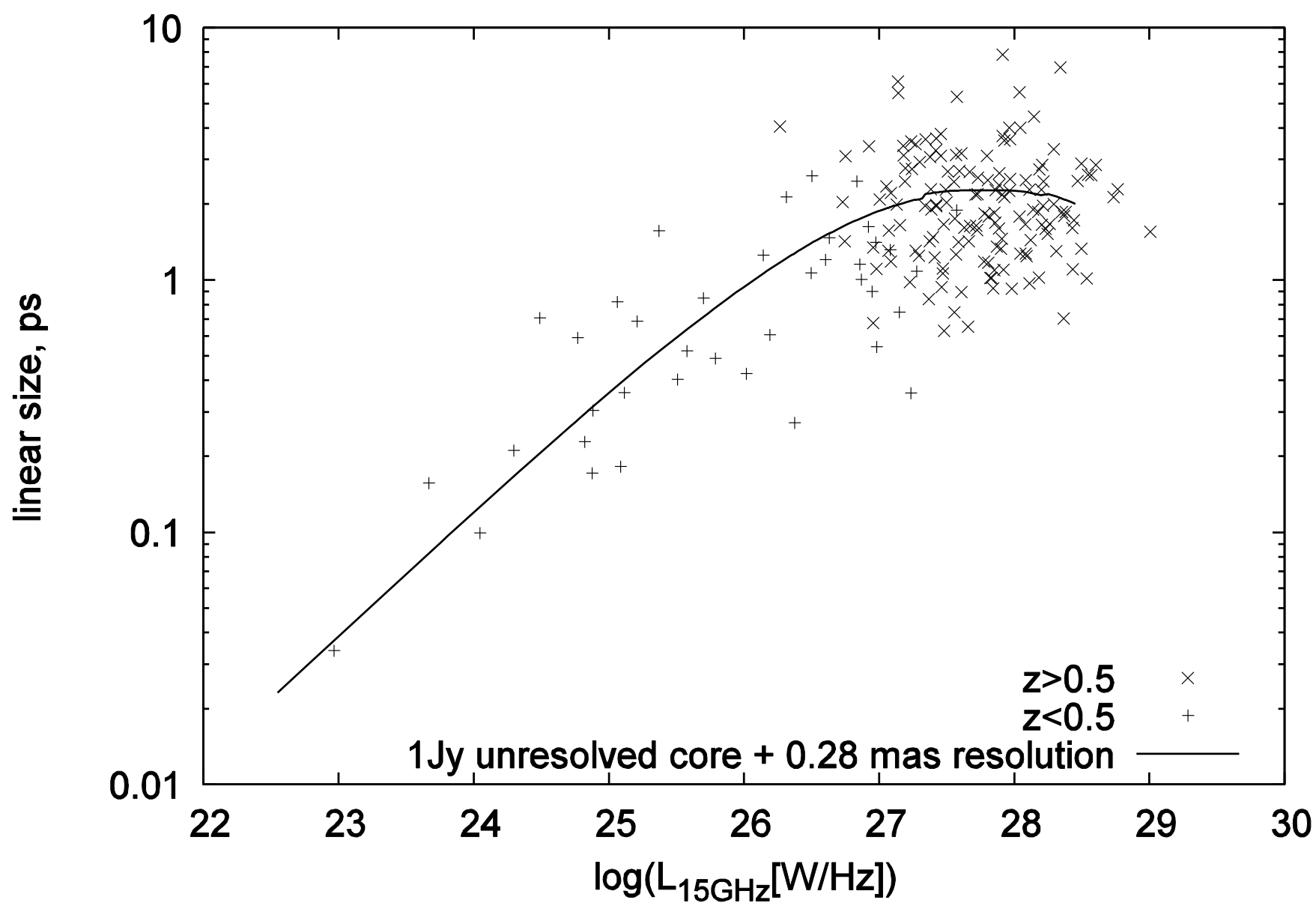

Fig.2 Linear size corresponding to core major axis vs. k-corrected 15 GHz VLBI-core luminosity for 186 sources where sources in different redshift ranges marked with different symbols

## 2. VLBI-CORE SIZE AS THE ANGULAR SCALE: 15 GHz DATA

To illustrate our statement we took a data from 15 GHz VLBA flux-limited sample (Kovalev et al., 2005) consisting of 250 flat-spectrum radio sources and used a major axis of VLBI-core as measure of the angular scale $\theta$. As the authors provide data for several epochs we used parameters on epoch where the unresolved flux density of the source $S_{unres}$ was greatest (see article for definition of $S_{unres}$ ). This increase in flux density could be due to birth of a new currently non-resolved component and confuse the measurements of the VLBI-core major axis (increase it), but we ignore this complication as it is hard to picture how it could affect our results. Among 197 sources with known redshifts 186 have estimates of core major axis size from modeling of the visibility function. We binned the data on $z$ in bins with near equal number of objects (~20) and used mean major axis in bins as the angular scale on the mean redshift in each bin. The "$\theta - z$" plot is presented in Figure 1 with model curve, representing a standard rod of a known linear size in the Universe with cosmological parameters $H_0$=70$km/(s \cdot Mpc)$, $\Omega_\Lambda$=0.7, $\Omega_M$=0.3 (that used in article unless otherwise stated). First of all it can be seen that point with $z$<0.5 lies under model curve. Practically, this obvious complication is avoided by just discarding that data. This is done through considering "luminosity-linear size" relation and finding correlation of luminosity with linear size on $z$<0.5 as mentioned above. Although some authors attribute this correlation to the intrinsic source properties (such as transition of sources from non-relativistic to ultra relativistic regime during which luminosity rises to a constant intrinsic value – Jackson, 2004) it seems that things are much simpler. On Figure 2 we plot linear sizes of respective major axis vs. k-corrected 15 GHz luminosities of VLBI-cores from 15 GHz data of (Kovalev et al., 2005). Also we plotted the curve, corresponding to 1 Jy VLBI-core of constant angular size which is taken to be 0.28 mas as indicated by fitting "$\theta - z$" relation (of course this is a simplification because resolution depends on uv-plane coverage obtained in observations of each single source). This curve models the unresolved VLBI-core component. As can be seen such correlation occurs because of
angular scale corresponding to the major axis of VLBI-core doesn't depend on redshift. In other words the measured quantities represent not any physical linear scale but the ever existing "structure" in AGN jets on the angular scale of ground-based VLBI resolution.

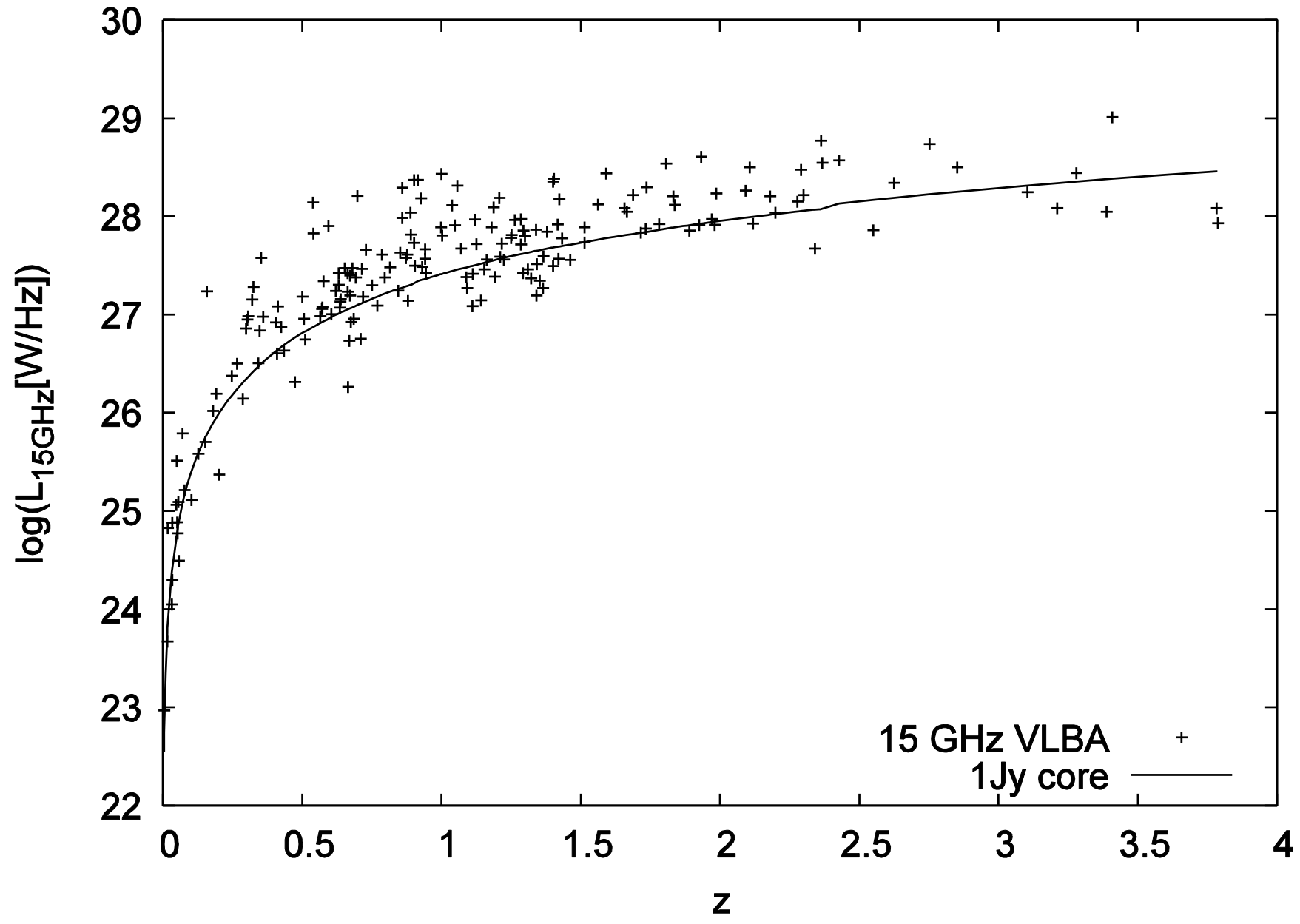

Fig.3 Luminosity – redshift plot demonstrating the Malmquist bias in our flux limited sample

Considering "luminosity-redshift" (Fig.3) and "linear size – redshift" (Fig.4) plots clarifies the issue. The first illustrates the Malmquist bias originating in any flux-limited sample. The redshift $z$ is the same as the luminosity $L$ in such samples (because of the steep radio luminosity function). Now it is clear why there is a "cloud" of points in the "luminosity-linear size" relation on $z$>0.5. In our data this cloud occupies region of $\log L[W/Hz]$=27÷29. From Figure 3 one can see that this range of luminosities corresponds to wide range of redshifts $z$=0.75-4 in our sample. In concordance cosmological model angular diameter distance weakly depends on redshift in this range of $z$. So all sources with $z$>0.75 "squeeze" in range $\log L[W/Hz]$=27÷29 on "luminosity – linear size" plot thus forming a "cloud". Of course the observed luminosities have nothing in common with intrinsic luminosities that could depend on parameters of central engine such as mass or spin of SMBH just because of Doppler boosting[3].

Thus the "standard linear" size of ~$3h^{-1}$ pc obtained by (Jackson, 2008) is just linear size corresponding to the resolution of VLBI network at $z$>0.5. It is nearly 3 times larger than one for our 15 GHz sample (as can be seen from Figure 1 our "standard rod" has a length of ~ $1h^{-1}$ pc ) consisting with ratios of observing frequencies (15 to 5 GHz) and different VLBI networks used (resolution achieved in VLBI experiments depends on uv-plane coverage that varies for different networks, source declination and duration of source tracking and the observing frequency).

[3] authors of (Cohen et al., 2007) give statistical estimate of the peak intrinsic luminosity of 15 GHz MOJAVE VLBA sample ~$10^{25÷26}$ $W/Hz$

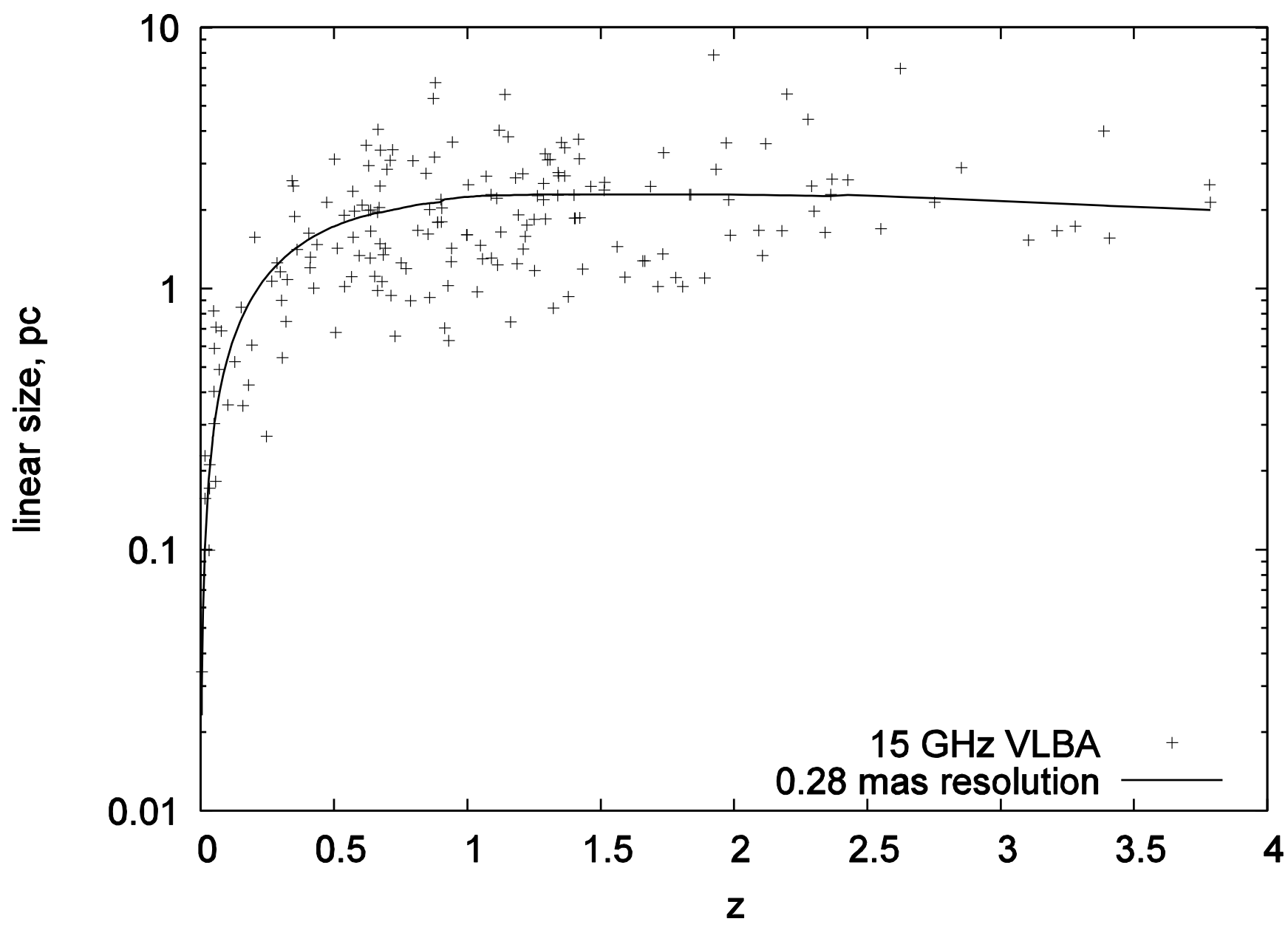

Fig.4 Linear size corresponding to VLBI-core major axis size vs. redshift. Solid line represents constant resolution of 0.28 mas

## 4. CHOISE OF THE ANGULAR SCALE MEASURE FOR ULTRA COMPACT RADIO SOURCES

Now it is clear that "luminosity - linear size" correlation for VLBI-cores has different origin than that for the same radio galaxies on the arcsecond scales and it represents a limited resolution of VLBI network. This is essential fact, provided that cores of VLBI jets often stay unresolved. Space baselines with HALCA satellite have demonstrated that some VLBI-cores observed on ground baselines possess a structure (Gabuzda, 2003). So until any resolution-independent scale (such as the surface of the "true" optically thick core) appears it makes sense to increase the resolution of VLBI-networks by extending baselines in space. And even after reaching the "true" core it remains unclear what exactly has to be done with the size of the core which depends on the emitted frequency (Blandford, Konigle, 1979) that is in turn subject to relativistic Doppler blueshift and cosmological redshift.

Also it is worth to note that if one defines the angular scale through compactness (ratio of visibility amplitudes in two points of uv-plane) the effect doesn't disappear. This suggests that the angular scale obtained through compactness is also determined not by some linear scale in the source structure but also by VLBI resolution. As jets can be represented as near self-similar structures from some outer scale down to core beyond which the emission is self-absorbed, they possess structures on every angular scale between. Thus using angular scale measures from compactness defined through core flux or visibility amplitude at sufficient spatial frequency one measures the resolution of VLBI network used in observations. In (Kovalev, Kardashev, 2000) plots of compactness parameters versus redshift are presented for 15 GHz VLBA data. The compactness defined through core flux $S_c$ to total flux $S_t$ on VLBA map or through core flux to single-dish flux $S_s$ does not depend on redshift, but compactness defined as the ratio of VLBA flux $S_{VLBA}$ to single dish flux increases with $z$ (in sense that there are no sources with high redshifts and small compactness or, equivalently, large angular sizes of VLBI jet on scales of tens of milliarcseconds). Implying that compactness is the measure of the characteristic angular size we can conclude that the size of VLBI-core doesn't depend on redshift, but thr size of VLBI jet does. The authors attribute last fact to "k-correction"- like the effect, described below.

In our opinion the most robust measure of angular scale for AGN jets observed on milliarcsecond scale is the separation between VLBI-core and some structure detail (component) in jet. This measure has been already used in works of Kellerman, 1993 (see also Gurvits et al., 1999). But even this choice that is free of resolution effects meets some systematic uncertainties (review of which could be found in Gurvits, 1994) such as frequency dependent shift of the core (Blandford, Konigle, 1979) and "k-correction" (Gurvits, 1994). Namely, emitted in the jet rest frame radiation is subject for Doppler blueshift because of relativistic bulk motion of the emitting matter and subsequent cosmological redshift. So if the observed frequency is $\nu_{obs}$ then the emitted one is $(1+z)\nu_{obs}/\delta$, where $z$ - redshift and $\delta$ - Doppler factor of jet bulk motion. As the separation of $\tau$=1 surface from optically thin features in VLBI jets increases with frequency and the jet emission has a steeper spectrum but the core has practically flat spectrum this would introduce some uncertain change in the observed jet length. Effect is even more dramatic if we compare beamed jet emission with extended unbeamed (that is captured in single dish flux) for which only cosmological redshift is at work. In fact the authors of (Kovalev, Kardashev, 2000) could have observed this case (see note above and discussion on the projection effects below). Although relativistic and cosmological frequency shifts work in different directions the fact of their cancellation is not quite certain (see APPENDIX A).

## 5. CONCLUSIONS

We have shown that ultra compact radio sources observed on scale of milliarcseconds with VLBI should be regarded as standard rods with caution. Existing data supports the conclusion that the sizes of VLBI-cores do not represent any physical scale in source but rather limited resolution of ground VLBI networks. Thus it is worth to increase the resolution of VLBI observations by extending the baselines in space until some stable physical linear scale is reached. Other way to avoid the interference of resolution effects is to consider distance between VLBI-core and some knot in jet as the angular scale (that was first done by Kellerman, 1993), but even this choice meets some difficulties because of quite uncertain values of Doppler blueshifting that has to be estimated somehow.

## ACKNOWLEDGMENTS

Authors thank the members of Astro Space Center scientific seminars for useful discussions.

## APPENDIX A. CANCELLATION OF COSMOLOGICAL AND DOPPLER FREQUENCY SHIFTS.

The observed flux of a radio source observed at redshift $z$ is:

$$F_\nu = \frac{\delta^p L_\nu^{int} (1+z)^{1-\alpha}}{4\pi d_L^2(z)} \qquad (1)$$

where ν is the frequency in observer rest frame, $L_\nu^{int}$ - intrinsic luminosity of jet (in jet rest frame), p=2+α for stationary photosphere (such as τ=1 surface), $p$=3+α for relativistically moving photosphere (such as moving "blobs"), α – spectral index ($F_\nu \sim \nu^{-\alpha}$), $d_L(z)$ – the luminosity distance to source. This expression differs from standard (definition of the luminosity distance) through Doppler boosting (that is described by $\delta^p$ - term) and cosmological k-correction (through $(1+z)^{1-\alpha}$ - term). For two sources to be visible at $z$=0.1 and $z$=2 with fluxes equal to the fixed survey flux limit the relation must be satisfied (using $H_0$=70$km/(s \cdot Mpc)$):

$$\left(\frac{\delta_{z=2}}{\delta_{z=0.1}}\right)^p \frac{L_{\nu,z=2}^{int}}{L_{\nu,z=0.1}^{int}} \approx 400 \qquad (2)$$

Taking p=2(or 3)+α, where α=0 (that represents VLBI-core which nearly always the brightest part of the source on VLBI-scale and consequently, its emission determines including of the source in flux-limited sample) one gets that without luminosity evolution:

$$\frac{\delta_{z=2}}{\delta_{z=0.1}} \sim 10 \text{ (11 for p=2.5)}, \qquad (3)$$

Where $\delta_z$ – Doppler factor needed for source with intrinsic luminosity $L_{\nu,z}^{int}$ and redshift $z$ to fall in flux limited sample. With some luminosity evolution $\frac{L_{\nu,z=2}^{int}}{L_{\nu,z=0.1}^{int}}$=10 (one should keep in mind that apparent absence of such evolution is one of the advantages of ultra compact radio sources in context of searching for standard rod) one obtains:

$$\frac{\delta_{z=2}}{\delta_{z=0.1}} \approx \text{ 4-7 (5 for } p\text{=2.5)}, \qquad (4)$$

The brightness temperatures dependence on $z$ also supports higher Doppler-factors of high redshift sources. If the observed brightness temperature undergoes Doppler enhancement then $T_{obs} = \delta T_{int}/(1+z)$, where $T_{obs}$, $T_{int}$ - observed and intrinsic (in jets rest frame) values. Using data from (Kovalev, Kardashev, 2000) and (Kovalev et al., 2005) (where quantities $T_{obs}(1+z) = \delta T_{int}$ are given) one can see that the highest observable temperatures are subject for more than an order of magnitude increase with redshift thus confirming the result that high-redshift sources are beamed more strongly.

Redshift dependence of superluminal velocities $\beta_{app}$ in VLBI jets could also be reconciled with higher Doppler-factors of distant sources. Although observational data points on a distribution of jet Lorentz factors $\gamma$ rather than universal value (Kellerman et al., 2004), high $\delta$ of distant sources (needed for including them in flux-limited sample) can only be achieved with high $\gamma$ of jets (see figure 20 in Urry, Padovani, 1995) and small angle to the line of site $\theta$. The most probable angle for sources in flux limited sample and Doppler bias is $\theta \sim (2\gamma)^{-1}$ where $\beta_{app} \sim \gamma/2$ – (Vermeulen, Cohen, 1994). The angle that maximize apparent velocity is $\sim\gamma^{-1}$ where $\beta_{app} \approx \gamma$ - (Urry, Padovani, 1995), so the maximum observable apparent speed (an upper envelope on $\beta_{app} - z$ plot) should increase with redshift, although there always should be “slow” sources. That is exactly what was observed (Zhang, Fan, 2007).

In any case discussed above the ratio of the emitted frequencies:

$$\frac{\nu_{z=0.1}^{emm}}{\nu_{z=2}^{emm}} = \frac{(1+0.1)\delta_{z=2}}{(1+2)\delta_{z=0.1}} > 1 \qquad (5)$$

~2 and ~4 for p=2.5 with and without supposed luminosity evolution. Thus opposite to the common wisdom, the jets of the sources on higher redshifts do not have higher rest frame frequencies of observed emission than nearby sources and do not need to be shorter because of their steeper spectra (via $k$-correction). But as discussed in (Jackson, 2004 and reference therein) in fact they are. As mentioned in that paper it could be so because of the projection effect (Dabrowski et al., 1995) – higher Doppler factors achieved at small angle of jet to the line of sight and the source must be seen nearly pointing towards us thus minimizing the jet projected length. Also at higher redshifts CMB energy density and inverse Compton losses increase and could heavily restrict observable jet length.